\documentclass[a4paper,11pt]{article}
\pdfoutput=1 

\usepackage{jcappub} 

\usepackage[T1]{fontenc} 

\usepackage{amssymb}
\usepackage{amsmath}
\usepackage{graphicx}
\usepackage{epsfig}
\usepackage{ccaption}
\usepackage{color}
\usepackage{fancyhdr}
\usepackage{textcomp}
\usepackage{tikz}
\newcommand{\be}{\begin{equation}}
\newcommand{\ee}{\end{equation}}
\newcommand{\bea}{\begin{eqnarray}}
\newcommand{\eea}{\end{eqnarray}}
\newcommand{\bes}{\begin{subequations}}
\newcommand{\ees}{\end{subequations}}

\newcommand{\bc}{\begin{center}}
\newcommand{\ec}{\end{center}}
\usepackage{wasysym}

\title{\boldmath Probing the Seesaw Mechanism with Cosmological Data}

\author[a]{J. G. Rodrigues,}
\author[b,c]{Micol Benetti,}
\author[d]{Marcela Campista,}
\author[d,a]{and Jailson Alcaniz}

\affiliation[a]{Departamento de F\'{\i}sica, Universidade Federal do Rio Grande do Norte, 59078-970,
Natal, RN, Brasil}
\affiliation[b]{Dipartimento di Fisica  ``E. Pancini", Universit\`a di Napoli  ``Federico II", Via Cinthia, I-80126, Napoli, Italy}

\affiliation[c]{Istituto Nazionale di Fisica Nucleare (INFN), sez. di Napoli, Via Cinthia 9, I-80126 Napoli, Italy}

\affiliation[d]{Departamento de Astronomia, Observat\'orio Nacional, Rio de Janeiro, RJ, Brasil}

\emailAdd{jamersoncg@gmail.com}
\emailAdd{micol.benetti@gmail.com}
\emailAdd{campista@on.br}
\emailAdd{alcaniz@on.br}

\abstract{
We investigate cosmological consequences of an inflationary model which incorporates a generic seesaw extension (types I and II) of the Standard Model of Particle Physics. A non-minimal coupling between the inflaton field and the Ricci scalar is considered as well as radiative corrections at one loop order. This connection between the inflationary dynamics with neutrino physics results in a predictive model whose observational viability is investigated in light of the current cosmic microwave background data, baryon acoustic oscillation observations and type Ia supernovae measurements. Our results show that the non-minimal coupled seesaw potential provides a good description of the observational data when radiative corrections are positive. Such result favours the type II seesaw mechanism over type I and may be an indication for physics beyond the Standard Model. }

\begin{document} 
\maketitle
\flushbottom

\section{Introduction}

The advances in our ability to map the cosmic history through the Large-Scale Structure (LSS) and Cosmic Microwave Background (CMB) observations have been fundamental to establish the standard $\Lambda$ - Cold Dark Matter ($\Lambda$CDM) model.  Requiring half a dozen parameters, this model is highly successful at explaining a number of observations under the assumptions that gravity is described by Einstein's general relativity, the spatial sections of the Universe at constant cosmological time are homogeneous and isotropic, and the existence of the dark matter and dark energy components. In this scenario, the evolution of the very early universe is described by the inflationary paradigm, which is able to provide an explanation for the origin of inhomogeneities in the Universe based on causal physics.


On the other hand, the Standard Model of fundamental particles (SM) -- based on the gauge group $SU(3)_C \otimes SU(2)_L \otimes U(1)_Y$ --  describes all known fundamental particles and their interactions, and constitutes one of the most successful theories ever developed. However, regardless of its predictive power, there are a number of physical processes that are not described or predicted by the SM as, for instance, the existence of weakly interacting and stable particles (dark matter) or the oscillatory behavior observed for active neutrinos coming from the Sun \cite{Patrignani:2016xqp}.  

These results motivated the proposal of a number of extensions of the SM. Among them, the Seesaw extensions furnish possible solutions to some of these problems \cite{Minkowski:1977sc,Yanagida:1979as,GellMann:1980vs,Mohapatra:1979ia,Schechter:1980gr}. Usually embedded in a gauge extension of the standard symmetries, the seesaw mechanism introduces a new scale of energy, associated with lepton number violation, in order to explain the smallness of neutrinos masses. This is accomplished through the introduction of new particles into the standard model content. The lightest stable particle, usually an axion or a sterile fermion, is often taken as a possible dark matter candidate.

In this work we connect the inflationary dynamics with neutrino physics through a single-field inflationary scenario. 
At low energy scale, the particle content of the seesaw mechanism is set, with new scalars and/or fermions added to the SM in order the generate neutrinos masses. We consider in our analysis prototype type I and type II seesaw mechanisms. At high energy, the scalar field responsible for the breaking of lepton number is free to rule inflation, and a nearly-flat inflationary potential is obtained through a non-minimal coupling between the inflaton and the Ricci scalar \cite{Salopek:1988qh,Starobinsky:1980te,Fakir:1990eg,Makino:1991sg}. As we are dealing with an inflaton field inserted in the seesaw scenario, the possible couplings between the inflaton and the particle content of the model play a key role on the inflationary dynamics. In particular, radiative corrections are supposed to furnish sizeable contribution to the inflationary potential. We analyse the observational viability of this non-minimal radiative inflationary model, namely SeeSaw Inflationary model (SI), in light of the most recent CMB, baryon acoustic oscillation and supernova data. As the main results, we place tight constraints on the parameters involved in radiative corrections and on the main cosmological parameters. We find out that CMB data clearly favour the type II seesaw mechanism in our prototype scenario.

\section{Seesaw Mechanism and Inflation} \label{Seesaw Sec}

In the past decades many variants of the seesaw mechanism were constructed  with different particle content and symmetries arrangements  \cite{Schechter:1980gr,Schechter:1981cv,Mohapatra:1986aw,Montero:2001ts,Freitas:2014fda,Asaka:2005an,Pires:2016dqq,Cogollo:2019mbd}. The type I and type II seesaw mechanism are among the most popular scenarios. In the most canonical case, type I seesaw adds three sterile fermions $N_{i}$ (right-handed neutrinos), $i=1,2,3$, and one complex scalar $\phi$, to the particle content of the standard model (SM). New symmetries are usually imposed to these fields in order to arrange them into the Yukawa terms, 
\be
\mathcal{L} \supset h_{ij}\bar{f}_{iL}i \sigma_{2}H^\dagger N_{j} +h^{\prime}_{ij} \phi \bar{N}^C_{i}N_{j}, \label{YuTermI}
\ee
where $f=(\nu \,\,,\,\, e)_L^T$ and $H=(H^+ \,\,,\,\, H^0)_L^T$ are the standard leptons and Higgs doublets, respectively. The electroweak symmetry breaking yields a Dirac mass component to standard neutrinos, while the second term in (\ref{YuTermI}) generate Majorana mass to the right-handed neutrinos. 
Once the Higgs field and the scalar singlet acquire vacuum expected values (vev), $v$ and $v_\phi$, $v \ll v_\phi$, tiny Majorana masses to standard neutrinos are generated. 

On the other hand, type II seesaw mechanism includes to the SM field content just a scalar, $\Delta$, triplet under $SU(2)_L$ SM symmetry \cite{Schechter:1980gr,Ma:1998dx,Hambye:2000ui,Freitas:2014fda},
\begin{equation}
\Delta\equiv \left(\begin{array}{cc}
\frac{\Delta^{+}}{\sqrt{2}} & \Delta^{++} \\ 
\Delta^{0} & \frac{-\Delta^{+}}{\sqrt{2}}
\end{array} \right). 
\end{equation}
There is no need to extend the gauge symmetry structure of the SM, $SU(3)_C \otimes SU(2)_L \otimes U(1)_Y$, once it permits the Yukawa term,
\be
\mathcal{L} \supset  h_{ij}\bar{f}^{c}_{i} i \sigma_{2}\Delta f_{j}. \label{YukawaII}
\ee
When the neutral component of $\Delta$ acquire vev the standard neutrinos obtain Majorana masses proportional to $v_\Delta$.


In common, all those extensions adds to the standard model new scalars associated with high energy physics, some of the main ingredients for slow-roll inflation. This motivated a series of papers relating neutrinos physics and inflation \cite{Okada:2011en,Boucenna:2014uma,Ferreira:2016uao,Ballesteros:2016euj,Ferreira:2017ynu,Rodrigues:2018jpv,Borah:2018rca}. Usually,  in these scenarios the couplings between neutrino fields and the inflaton provide a contribution to the inflationary potential through radiative corrections. This contribution changes the shape of the inflationary potential and, consequently, the predictions to the inflationary parameters. 

The one-loop Coleman-Weimberg radiative corrections to the inflationary potential can be written in the form \cite{Coleman:1973jx},
\begin{equation}
 V(\Phi) \approx \frac{ \lambda}{4}\Phi^{4} + a\Phi^4 \ln{\frac{\Phi}{M}}, \label{potaprox}
\end{equation}
where $\Phi$ represents the inflaton field, $M$ is the renormalization scale, and $a$ encodes the radiative contributions to the inflationary potential. In the case of type I seesaw mechanism the natural candidate for inflaton is the real component of the complex scalar, $Re(\phi)$. The Yukawa interactions of the inflaton field contribute to the inflationary potential via loops of right-handed neutrinos,
\be
a \simeq \frac{-4{h^{\prime}}^4}{32\pi^2}, \label{TypeIradiative}
\ee
where ${h^\prime}^4 = \sum_i {h^\prime_i}^4$ and for simplicity we have used the basis which makes $h^\prime_{ij}$ diagonal. For type II seesaw mechanism the neutral component of the scalar triplet, $\Delta^0$, is the natural choice for inflaton candidate. As we are dealing with a weak triplet field, radiative corrections from the gauge sector, as well form Yukawa sector, must be considered. The dominant one-loop terms can be written in the form
\be
a \simeq \frac{3(g^4 +g^{\prime 4}) - 4h^{4}}{32\pi^2}, \label{TypeIIradiative}
\ee
where $g$ and $g^\prime$ are the standard $SU(2)_L$ and $U(1)_Y$ gauge couplings, respectively. Scalar interactions between the inflaton candidate and the Higgs field are also supposed to generate radiative corrections, although the measured Higgs mass imposes tight constraints on these couplings.

Although radiative corrections may contribute to the inflationary potential, it is well established that such corrections are not enough to align the model predictions with the observed values to inflationary parameters \cite{Martin:2013tda}. In particular, the Coleman-Weinberg potential (\ref{potaprox}) predictions to $n_S$ and $r$ are out of the $95\%$ C.L. region obtained by the Planck collaboration \cite{Aghanim:2018eyx}.

\section{Non-Minimal Inflation and the Unitarity Problem}

Theories of modified gravity have become popular in the past years, although their original application to the dynamics of the primordial universe dates back to the eighties~\cite{Starobinsky:1980te}.  
In general, modified gravity inflationary models predict tiny tensor-to-scalar ratio \cite{Salopek:1988qh,Starobinsky:1980te,Fakir:1990eg,Makino:1991sg}, $r$, in good agreement with current CMB observations \cite{Aghanim:2018eyx,Aghanim:2019ame}. In particular, the possibility of a non-minimal coupling of the inflaton with gravity has been widely debated after the proposal of a Higgs field driven inflation \cite{Bezrukov:2007ep,Barvinsky:2008ia,GarciaBellido:2008ab,Bezrukov:2014bra,Lee:2018esk}.

The non-minimal coupling of the inflaton field and the Ricci scalar is defined in the Jordan Frame. In general terms, the Lagrangian of the model is written in the form
\be
 {\cal L} = \frac{1}{2} (\partial_\mu \Phi)^{\dagger}(\partial^\mu \Phi)-\frac{M_P^2R}{2}-\frac{1}{2}\xi {\Phi}^2 R -V_J(\Phi),
 \label{Ljordan}
\ee
where $M_P=2.435\times 10^{18}$ GeV is the reduced Planck mass. In order to calculate the inflationary parameters, 
we perform the following conformal transformation 
\cite{Birrell:1982ix,Accioly:1993kc,Faraoni:1998qx}:
\be
\tilde g_{\mu \nu}=\Omega^2g_{\mu \nu}\,\,\,\,\,\,\mbox{where} \,\,\,\,\,  \Omega^2= 1+\frac{\xi {\Phi}^2}{M^2_P}.
\label{conformaltransf}
\ee
which makes the kinect energy of inflaton field non-canonical. The process is finished by the field redefinition
\be
\chi^\prime \equiv \frac{d\chi}{d\Phi}=\sqrt{\frac{\Omega^2 +6\xi^2 {\Phi}^2/M_P^2}{\Omega^4}}.\label{InflatonRed}
\ee
Finally, the Lagrangian with minimal gravity sector and canonical kinetic term is defined in the Einstein frame,
\be
 {\cal L} = -\frac{M^2_{P} \tilde R}{2}+\frac{1}{2} (\partial_\mu \chi)^{\dagger}(\partial^\mu \chi)-V(\chi)\,,
 \label{LEinstein}
\ee
where $V(\chi)=\frac{1}{\Omega^4}V_J(\Phi[\chi])$. Note that the descriptions in both frames agree in the weak-field regime \cite{Faraoni:1998qx}.

The Einstein frame potential in (\ref{LEinstein}) is the one employed to calculate the predictions for inflationary parameters. In particular, the Higgs inflation predictions of $n_S\simeq 0.97$ and $r \simeq 0.0033$ are in excellent agreement with the latest results of Planck2018 \cite{Aghanim:2018eyx}. Nevertheless, there is a price to pay. In order to reconcile the tiny observed value of the amplitude of scalar perturbations, $A_S \simeq 2.1 \times 10^{-9}$, with the electroweak Higgs field quartic coupling, $\lambda \sim 0.1$, one has to assume a huge non-minimal coupling $\xi \sim 10^4$. Far more dangerous than just an unnatural value, there are also claims that a non-minimal coupling of such magnitude gives rise to a new scale of unitarity loss \cite{Burgess:2009ea,Barbon:2009ya,Burgess:2010zq,Hertzberg:2010dc}. See \cite{Lerner:2009na,Bezrukov:2013fka,Rubio:2018ogq} for a different point of view.




In order to circumvent this problem we will consider in our analysis extensions of the standard model in which the inflaton field is only weakly and moderately coupled to the Ricci scalar, $0.1\leq \xi \leq 100$. As will be shown later, it raises the unitarity scale $\Lambda$ to a energy regime much higher than the inflationary energy.

\section{Slow-Roll Analysis} \label{SR Analysis}

In this section, we aim to investigate the non-minimal inflationary scenario in light of the seesaw paradigm, namely, the SeeSaw Inflationary model. Radiative corrections are supposed to contribute in the inflationary potential \cite{Barvinsky:1998rn,Barvinsky:2008ia,GarciaBellido:2008ab,Bezrukov:2008ej,Lerner:2009xg,Okada:2010jf,Hamada:2014iga,Bezrukov:2017dyv,Bostan:2019fvk}. 
There are two different procedures to compute loop corrections in non-minimal models, namely prescription I and prescription II. The difference resides in which frame the loops are evaluated. In this work we opt to employ prescription II. However, at one loop level both procedures are equivalent \cite{Masina:2018ejw}.

For prescription II, radiative corrections are computed in Jordan frame. Feynman rules reduce to its trivial form and the effective potential can be written in the form (\ref{potaprox}). After the conformal transformation, we have for the scalar potential in Einstein frame
\begin{equation}
\label{eq:EinPot}
 V(\Phi) \approx \frac{ \frac{ \lambda}{4}\Phi^{4} + a\Phi^4 \ln{\frac{\Phi}{M_P}}}{\left(1 + \xi \frac{\Phi^2}{M^2_P}\right)^2},
\end{equation}
where $\Phi$ is the non-canonical scalar field and we take $M_P$ as the renormalization scale.

The slow-roll parameters can be written as
\begin{eqnarray}
 &\epsilon & = \frac{M^2_{P}}{2}\left(\frac{ V^{\prime}}{ V \chi^{\prime}}\right)^2, \quad \quad
 \eta  = M^2_{P}\left( \frac{V^{\prime \prime}}{V \chi^{\prime}} - \frac{V^{\prime} \chi^{\prime \prime}}
 {V {\chi^{\prime}}^3}\right),
 \label{slowparameters}
\end{eqnarray}
where $^\prime$ indicates derivative with respect to $\Phi$. Inflation starts when
$\epsilon,\eta \ll 1$ and stops when $\epsilon,\eta = 1$. In the slow-roll regime, the spectral index and the
scalar-to-tensor ratio have the form,
\be
 n_S=1-6\epsilon+2\eta, \quad \quad \quad r=16\epsilon.
 \label{PParameters}
\ee

%
The amplitude of scalar perturbations is given by
\begin{equation}
A_S =\left.\frac{V(\Phi)}{24M^4_P\pi^2\epsilon(\Phi)}\right|_{k=k_{*}} ,
\label{eq:PR}
\end{equation}
where $_*$ refers to pivot scale, i.e., when the CMB modes exit horizon at the scale $\Phi_{*}$. The value of $A_S$ is set by the COBE normalization to about $2.1 \times 10^{-9}$ for the pivot choice $k_{*}=0.05$ Mpc$^{-1}$ \cite{Aghanim:2018eyx}. Now, by inverting the eq.~(\ref{eq:PR}), we can write the value of the amplitude $\lambda$. 
Noteworthy is the strict dependence of $\lambda$ with both the $\xi$ and $a$ parameters, or rather the degeneracy of such parameters in the value of the potential amplitude. 

\begin{figure}[!t]
\centering
\includegraphics[width=15cm,height=7cm]{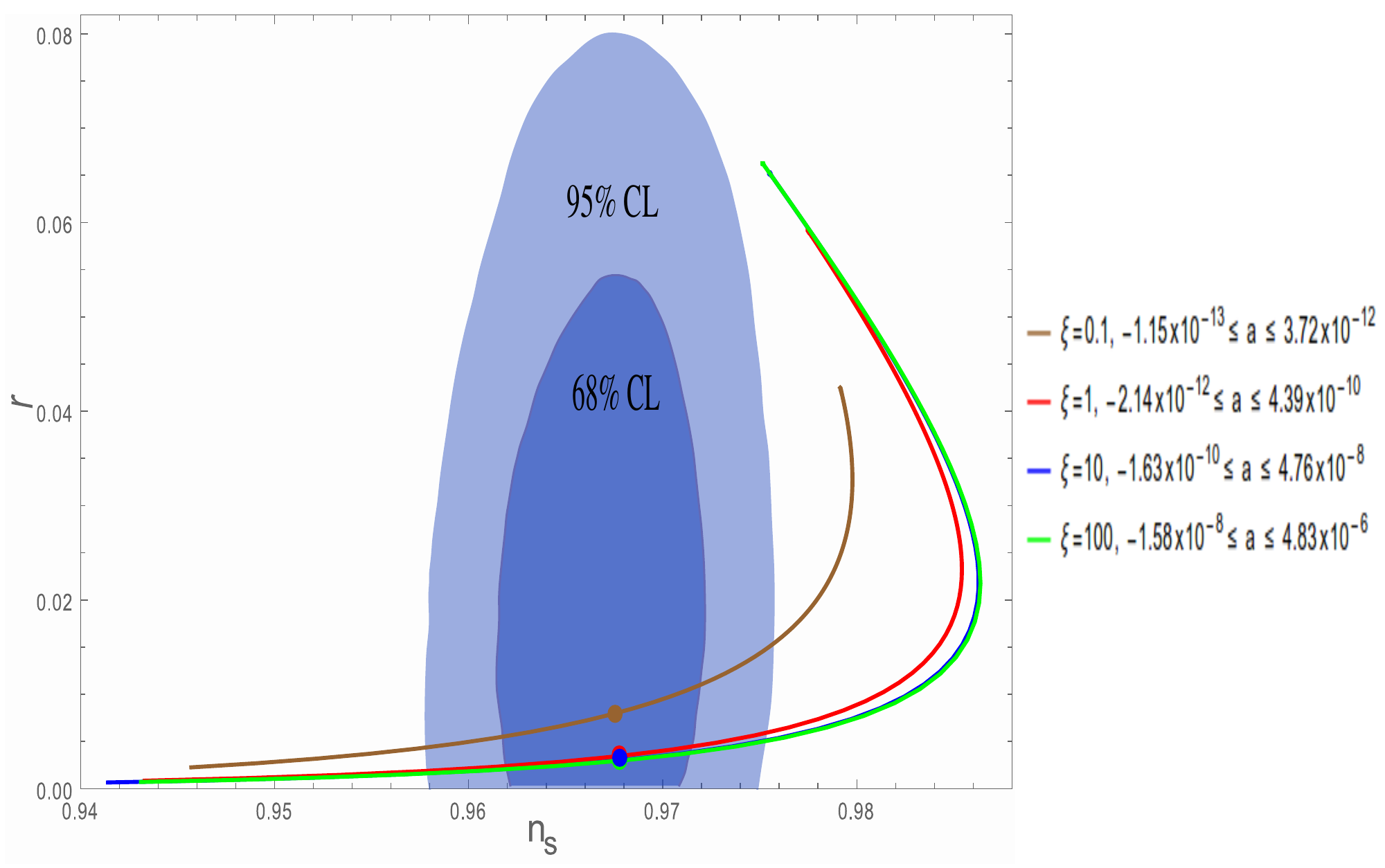}
\caption{$n_S$ vs $r$ for $\xi = 0.1$, $1$, $10$ and $100$. The grey areas show the favoured regions by Planck2018, with $68\%$ and $95\%$ confidence level (Planck $TT,TE,EE+lowE+lensing+BK14+BAO$ data set) \cite{Aghanim:2018eyx}.
}
\label{Fig:ns_r}
\end{figure}

The observable field value $\Phi_{\ast}$ can be related to the number of e-folds at which the pivot scale crossed out the Hubble radius during inflation, defined as 
\begin{equation}
    N_{\ast} = -\frac{1}{M^2_P}\int_{\Phi_{\ast}}^{\Phi_e}\frac{V}{V^\prime}{\chi^\prime} ^2d\Phi. \label{efolds}
\end{equation}
For the range of $\xi$ we are considering, $0.1\leq \xi \leq 100$, inflation ends as a $\Phi^4$ potential and the universe expands as radiation dominated in the reheating era \cite{GarciaBellido:2008ab,Bezrukov:2008ut,Repond:2016sol,Almeida:2018oid,Gong:2015qha}. This case is particularly interesting because it presents a reduced uncertainty in the estimation of the e-fold number. It is therefore plausible to affirm 
that in our model $N$ should be around 60 \cite{Liddle:2003as,Ballesteros:2016xej}. We then set $N=60$ in order to estimate the field strength in the horizon crossing, and use the resulting $\Phi$ to evaluate the inflationary parameters $n_S$ and $r$. In figure \ref{Fig:ns_r} we show our results in the $n_S \times r$ plane.

In order to draw the lines in figure \ref{Fig:ns_r} we vary  the parameter $a$ for fixed values of $\xi$, as depicted in the figure. 
In each curve we mark the points where $a=0$. The corresponding values for $n_S$ and $r$ indicate  that the Planck results are compatible with null radiative correction. Note also that the curves almost overlaps for $\xi \gtrsim 1$. This suggests that the inflationary observable become insensitive to variations in $\xi$ in the strong coupling regime. In fact, the predictions of Higgs Inflation of $n_S\simeq 0.968$ and $r \simeq 0.0033$ are practically identical to the case where $\xi \sim 100$ and $a=0$.

\begin{figure}[!t]
\centering
\includegraphics[scale=0.27]{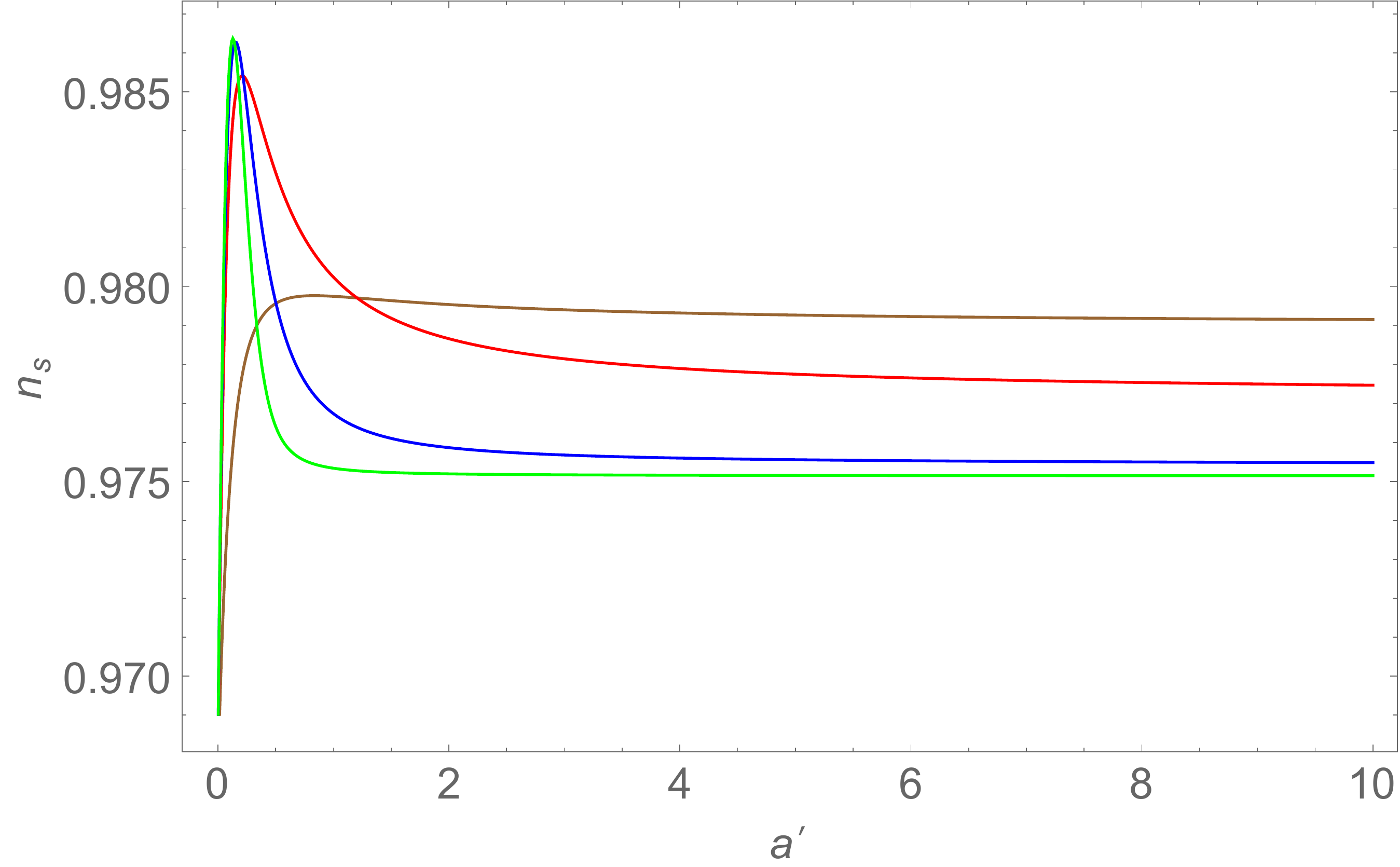}
\includegraphics[scale=0.27]{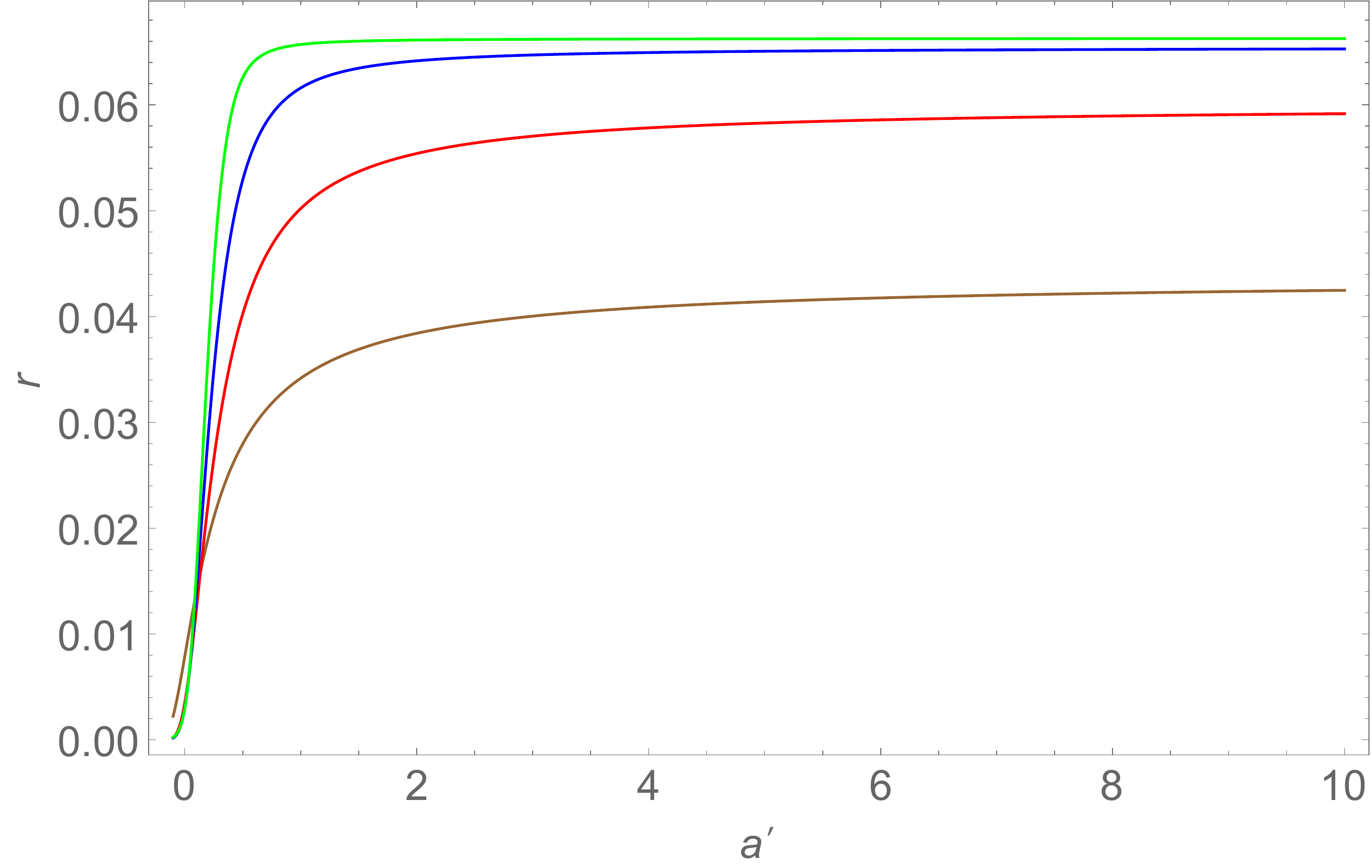}
\caption{$n_S$ vs $a^\prime$ (left) and $r$ vs $a^\prime$ (right) for $\xi = 0.1$, $1$, $10$ and $100$. In the figures $-0.1 \leq a^\prime \leq 10$.
}
\label{nsandrversusalinha}
\end{figure}

In order to evaluate the dependence of the inflationary parameters ($n_S$ and $r$) on the radiative corrections we make a choice of a suitable basis of independent variables. This makes necessary because the amplitude of scalar perturbation $A_S$, as measured by the Planck Collaboration in its last release, $2.101\times 10^{-9}$, constrains simultaneously $\lambda$, $a$ and $\xi$. For simplicity, we  define 
\begin{equation}
\label{eq:alinha}
a^\prime=4\frac{a}{\lambda}\;
\end{equation}
and $\xi$ as independent variables. The results are shown in  figure \ref{nsandrversusalinha}. Note that either $n_S$ and $r$ approach a constant value for sufficiently high values of $a^\prime$. The explanation for this behavior comes from the fact that for $a^\prime \gg 1$ the radiative correction $a$ overcomes the tree level contribution $\lambda$. In this regime $A_S$ constrains $a$ to its maximum value and $\xi$ is left as the only independent variable. 
These superior limits for radiative corrections are presented in figure \ref{Fig:ns_r} as the upper limits attributed to $a$ for each value of $\xi$. On the other hand, in the small regime of $a^\prime$ both $n_S$ and $r$ approach small values. After reaching a maximum, $n_S$ quickly becomes incompatible with the Planck favoured region. Particularly, a $\xi$-dependent minimum value for $a^\prime$, and consequently for $a$, is dictated by the e-folds number. However, as these values are highly incompatible with the Planck results for the $n_S$ parameter we disregard them
in our analysis.

\begin{figure}[!t]
\centering
\includegraphics[scale=0.33]{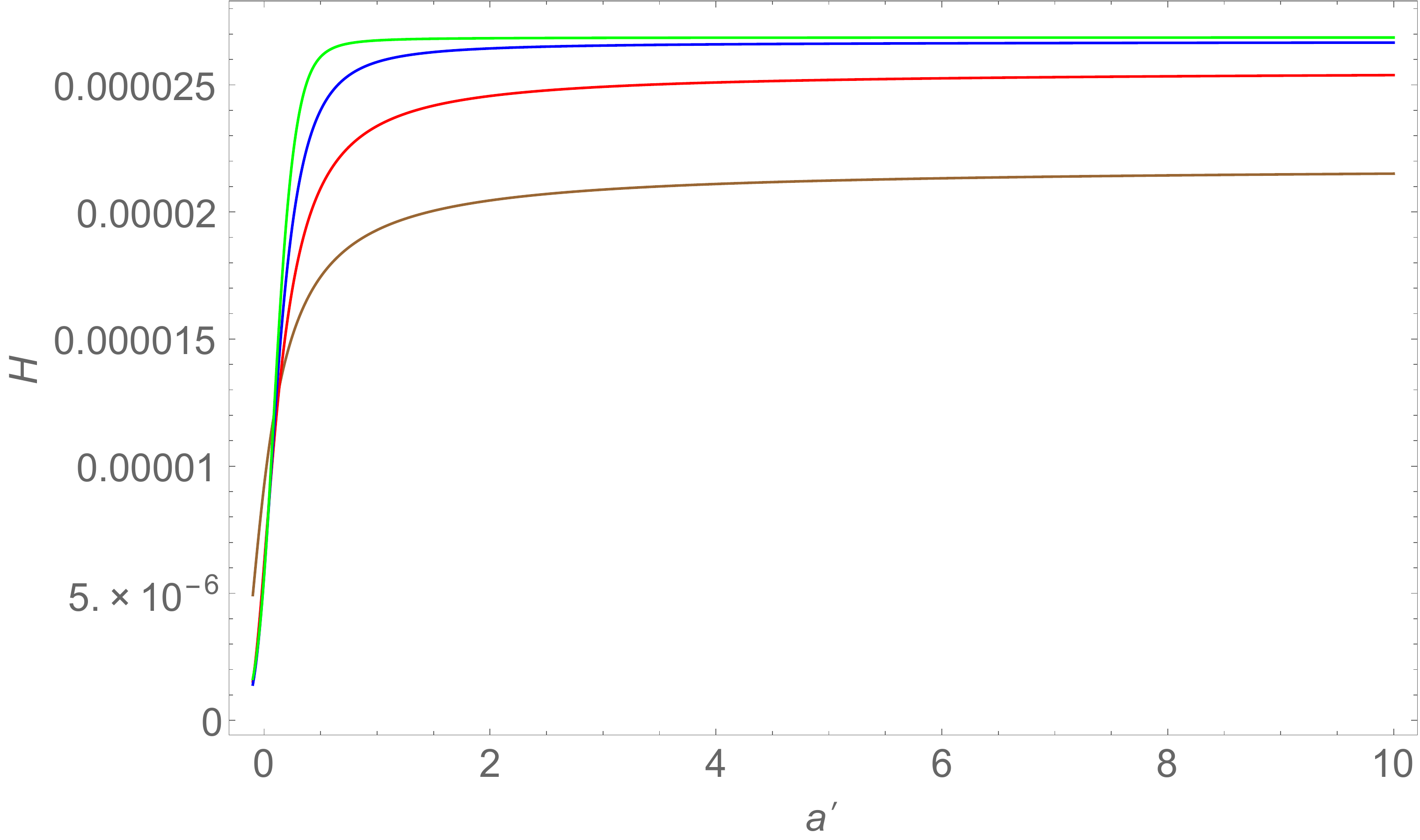}
\caption{$H$ vs $a^\prime$ for $\xi = 0.1$, $1$, $10$ and $100$. Colors are labeled as in Fig. \ref{Fig:ns_r}.
}
\label{EnergyScale}
\end{figure}

The scale of energy at which inflation takes place, $k^0 \sim H$, is depicted in figure \ref{EnergyScale}. Here we show our results in terms of Planck units, $M_P=1$. Notice that even for the highest value of $\xi$ considered, the unitarity scale $\Lambda = 1/\xi$ is orders of magnitude above the inflationary energy scale. Therefore, it is reasonable to say that the model discussed here is unitarily safe.

\begin{figure}[!t]
\centering
\includegraphics[scale=0.28]{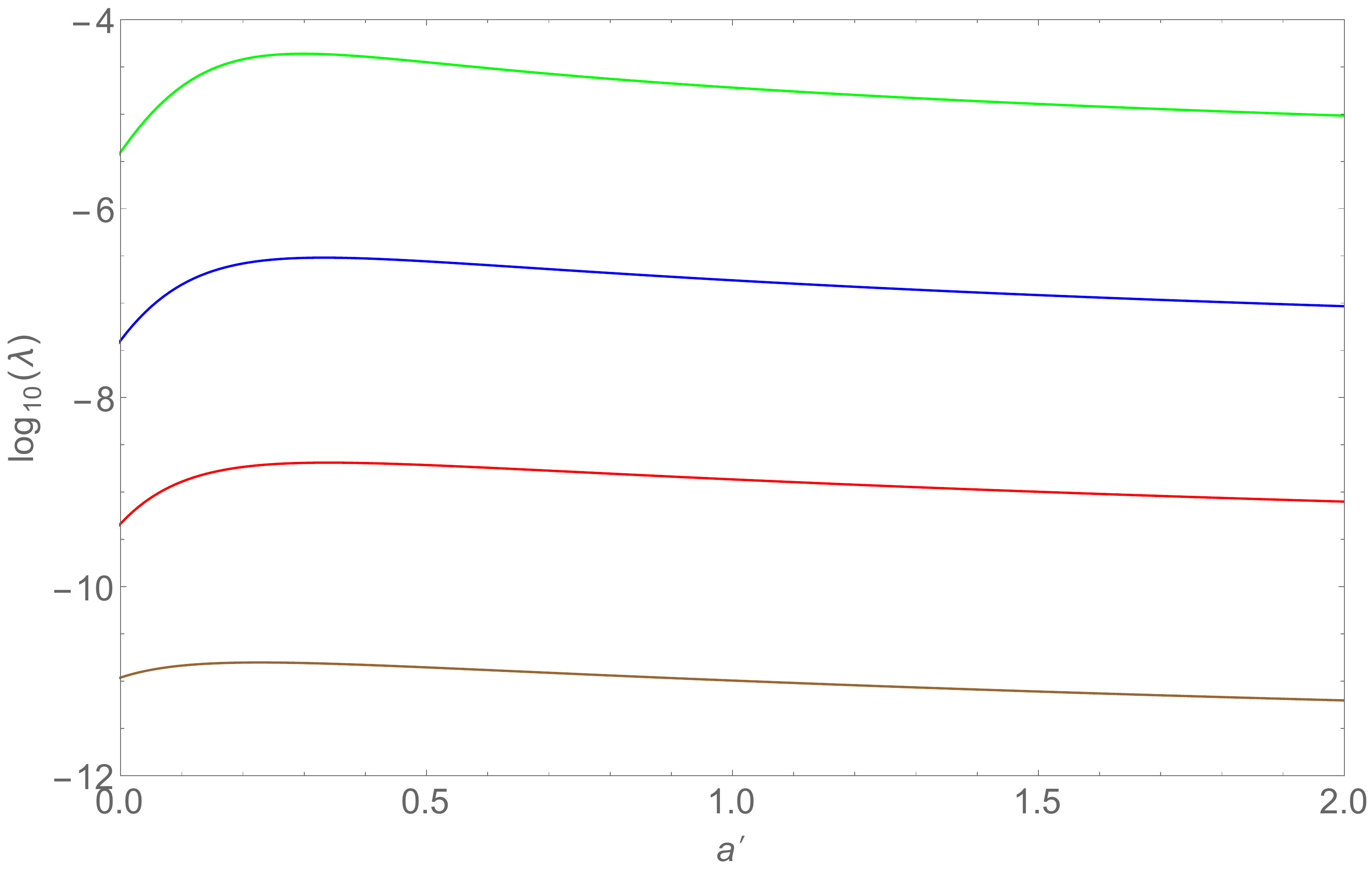}
\caption{$\log(\lambda)$ vs $a^\prime$ for $\xi = 0.1$, $1$, $10$ and $100$. Colors are labeled as in Fig. \ref{Fig:ns_r}.
}
\label{lambversusalinha}
\end{figure}

Finally,  figure \ref{lambversusalinha} shows the dependence of $\lambda$ with $a^\prime$ for fixed values of $\xi$. As mentioned above, the correlation between these three parameters depends on $A_S$. In particular, note that for each curve $\lambda$ reaches a maximum value somewhere between $0.1 \leq a^\prime \leq 0.3$. This maximum becomes more evident for higher values of $\xi$. As we shall see on the next section, this behavior will be extremely important for constraining $a^\prime$.

\section{Method}
\label{Method}
In this section we introduce the codes used in our analysis and the sets of  observational data used to perform the parameters estimation.

In order to resolve the Boltzmann equations and compute the theoretical predictions of the SeeSaw Inflationary model, among which the CMB temperature power spectrum, we modified the current version of the Code for Anisotropies in the Microwave Background ({\sc CAMB})~\cite{Lewis:1999bs}, so that the primordial spectrum\footnote{In the standard {\sc CAMB} code, the primordial power spectrum is parameterized as the power-law $P_R=A_s {\frac{k}{k_{\ast}}}^{(n_s -1)}$. }
is now led by the inflationary potential of eq. (\ref{eq:EinPot}). Such a modification is performed following the lines of~{\sc ModeCode}~\cite{Mortonson:2010er, Easther:2011yq}, where the  CMB anisotropies spectrum is calculated solving numerically the inflationary mode equations, i.e., solving the Friedmann and Klein-Gordon equations as well as the
Fourier components of the gauge-invariant quantity $u$ for an exact form of the single field inflaton potential.
By integrating these equations it is possible to obtain $H$ and $\chi$ as a function of time and the solution $u_k$ for the mode $k$. Therefore, following these steps, the code can compute the power spectrum of the curvature perturbation $P_\mathcal{R}$ by $P_\mathcal{R} = \frac{k^3}{2\pi^2}\left|\frac{u_k}{z}\right|^2$, evaluated when the mode crosses the horizon \cite{Easther:2011yq},  with $z \equiv \dot{\chi}/H$(dots denote derivatives with respect to conformal time).
Furthermore, parameters estimation is obtained using a Monte Carlo Markov chain statistical analysis, modifying the available parameter estimation packages {\sc CosmoMC}~\cite{Lewis:2002ah} to our purpose.

\begin{figure}[!t]
\centering
\includegraphics[scale=1]{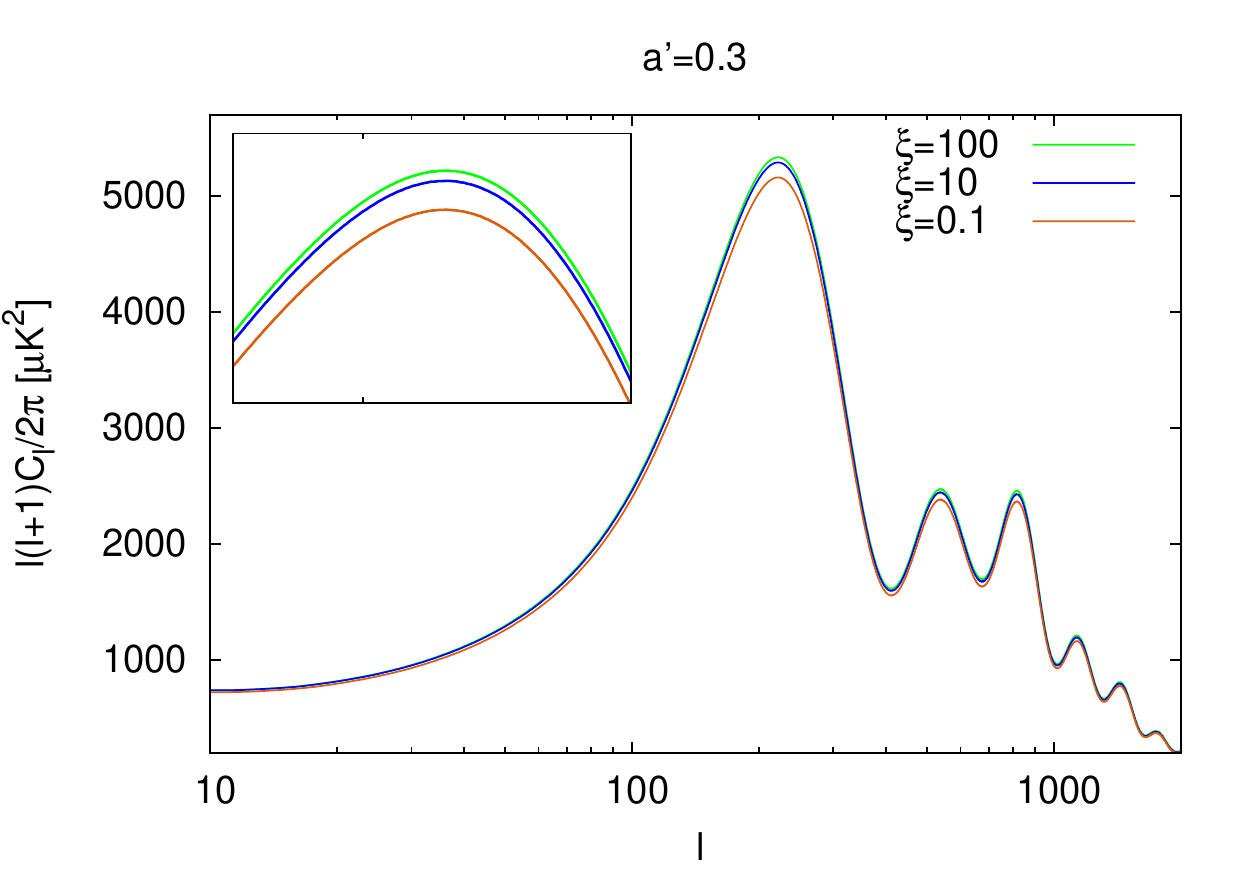}
\caption{Temperature power spectrum predictions for the SeeSaw inflation model fixing $a\prime=0.3$ and varying the coupling parameter value. In the box, a zoom of the first anisotropy peak at $\ell \sim 200$.}
\label{Fig:TT_alinha}
\end{figure}
\begin{figure}[!t]
\centering
\includegraphics[scale=1]{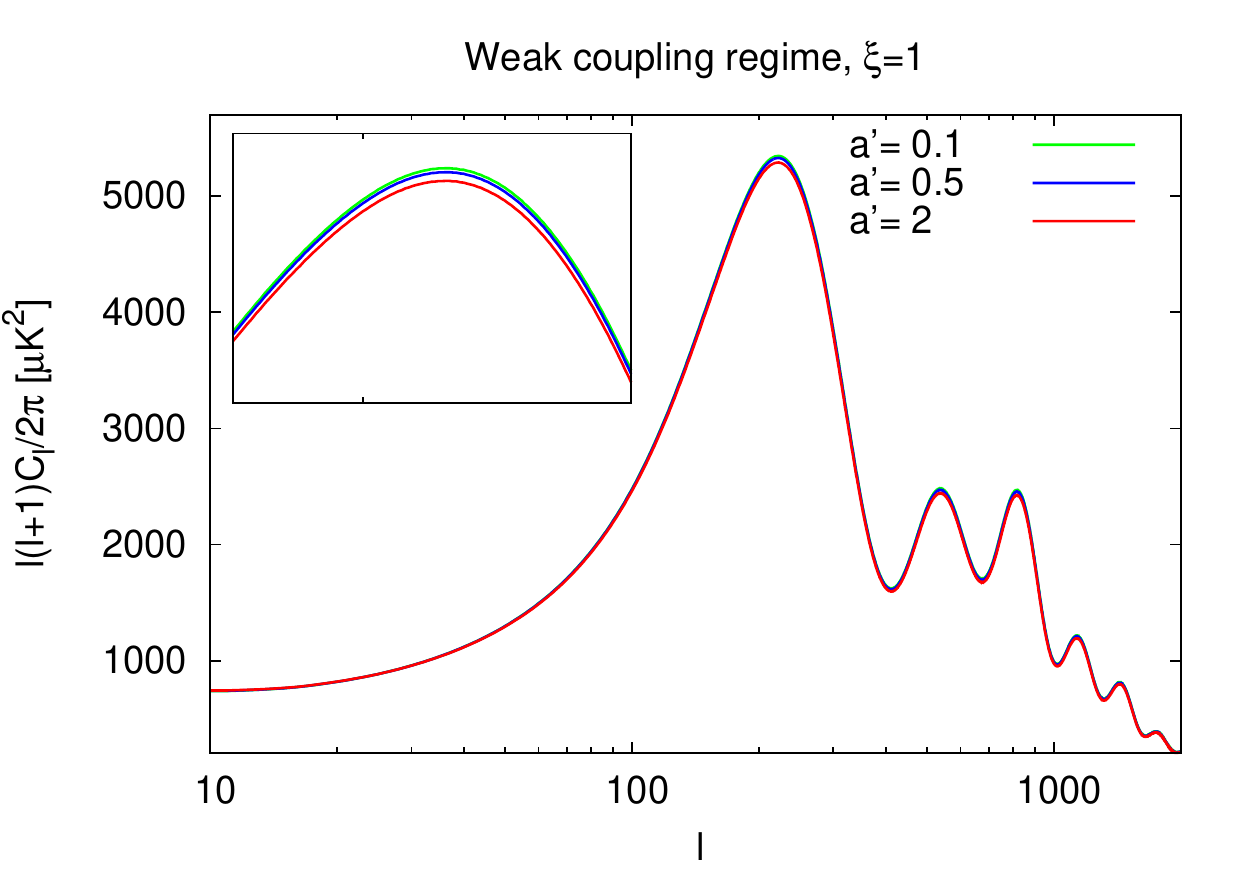}
\includegraphics[scale=1]{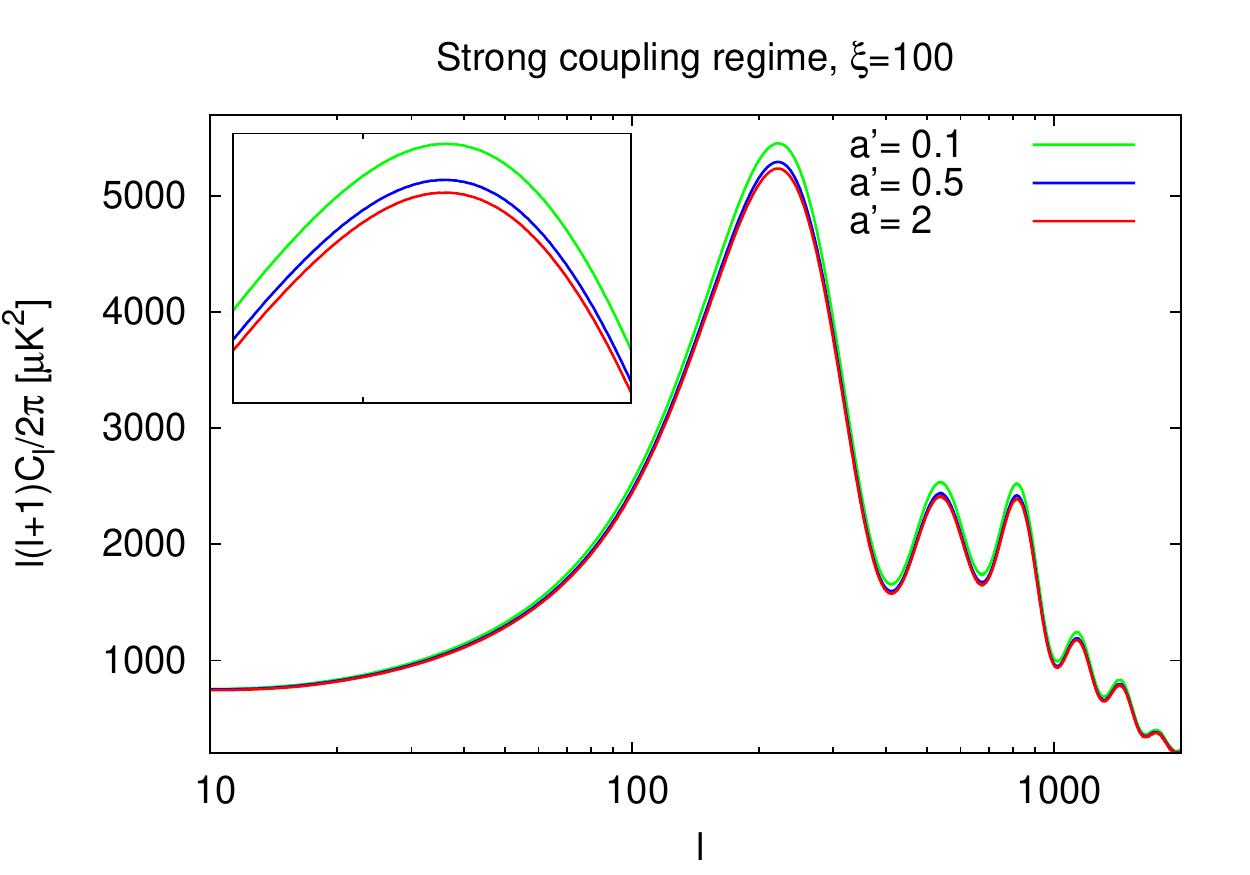}
\caption{Temperature power spectrum predictions for the SeeSaw inflation model fixing $\xi=1$ (top panel) and $\xi=100$ (bottom panel) and varying the $a\prime$ parameter value. In the boxes, a zoom of the first anisotropy peak at $\ell \sim 200$.}
\label{Fig:TT_xi}
\end{figure}

We perform our preliminary analysis assuming large flat priors on the cosmological parameters, while the values of the parameters $\xi$ and $a^\prime$ are chosen from the considerations made in the previous section, which we summarize below.
First, let us emphasize that the tensor-to-scalar ratio proves to be weakly sensitive to variations of $\xi$ (see figure \ref{Fig:ns_r}) for values of $n_S$ at $1\sigma$ C.L., while in figure \ref{nsandrversusalinha} we can note that both $n_S$ and $r$ show the same predictions for $a^\prime > 2$, i.e. they do not produce variations in the observables and, therefore, are not distinguishable from the data. Moreover, a very important variable to consider in order to select the inflationary parameters priors is the amplitude, $\lambda$ \cite{Campista:2017ovq, Santos:2017alg}. Let us recall that $\xi $ and $a^\prime$ parameters enter in the amplitude relation -- see eq. (\ref{eq:PR}) -- and consequently affect the amplitude of the temperature power spectrum. This observable is constrained with great accuracy by the observations of CMB and turns out to be the main observable for constraining the free parameters of the inflationary model. 
Looking at figure \ref{lambversusalinha}, it is clear that higher amplitude values are mainly achieved in the range $0.1<a^\prime<0.3$ for any fixed values of $\xi$, and that strong coupling regimes allow for larger $\lambda$. 
Unfortunately, the differences in $\lambda$ seen above between weak coupling regime (i.e. $\xi=1$ ) and strong coupling regime (i.e $\xi=100$) are very small for the different theoretical predictions of the temperature CMB power spectrum, as shown in figure \ref{Fig:TT_alinha}, making this parameter less performing than hoped for, but still the most sensitive to changes in inflationary parameters (compared to $n_s$ and $r$). Analyzing the model in weak coupling regime -- see figure \ref{Fig:TT_xi} top panel -- it does not appear to be an interesting option, given the negligible sensitivity (inside the error bar of the most recent CMB data) of the $a^\prime$ parameter on the observed amplitude (we consider up to $a^\prime=2$, since for higher values all theoretical predictions become insensitive to variations in $a^\prime$).
On the other hand, analysing the strong coupling regime (bottom panel of figure \ref{Fig:TT_xi}), it allows for higher amplitude variation range.
Thus, considering what has been mentioned above, we choose to analyse the strong coupling regime with $\xi$ fixed at $100$, assuming a flat prior for $a^\prime$ in the range $[-0.2:2]$.

For our purpose we consider two models, namely, the standard $\Lambda$CDM scenario (as the  model) with one parameter extension, i.e., leaving the tensor-to-scalar ratio as free parameter, and the Non-Minimal Radiative inflationary model varying the $a^\prime=4\frac{a}{\lambda}$ parameter.
Also, we vary the usual cosmological parameters, namely, the physical baryon density, $\Omega_bh^2$, the physical cold dark matter density, $\Omega_ch^2$, the ratio between the sound horizon and the angular diameter distance at decoupling, $\theta$ and the optical depth, $\tau$. In addition, we consider the nuisance foreground parameters~\cite{Aghanim:2015xee} and assume purely adiabatic initial conditions. The sum of neutrino masses is fixed to $0.06$ eV, and we limit the analysis to scalar perturbations with $k_{\ast}=0.05$ $\rm{Mpc}^{-1}$, also setting the arbitrary value of the number of e-folds $N_{\ast}=60$.

We choose to build two data sets for our analysis:
\begin{itemize}

\item Cosmic Microwave Background measurements, through the Planck (2018) data \cite{Aghanim:2019ame}, using temperature power spectra (TT) over the range $\ell \in [2 - 2508]$, and HFI polarization EE likelihood at $\ell \leq 29$;
\item BICEP2 and Keck Array experiments (BK15) B-mode polarization data \cite{Ade:2018gkx}, 
that improve the upper limit on the tensor-to-scalar ratio with respect to the previous release \cite{Array:2016afx}, constraining $r<0.07$;

\item Baryon Acoustic Oscillation (BAO): we use  distance measurements from 6dFGS ~\cite{Beutler:2011hx}, SDSS-MGS~\cite{Ross:2014qpa}, and 
BOSS DR12~\cite{Alam:2016hwk} surveys, as considered by the Planck collaboration;

\item Supernovae Type Ia (SNe Ia): We use the Pantheon sample that is the latest compilation of 1048 data points, covering the redshift range $[0.01 : 2.3]$~\cite{Scolnic:2017caz}.
\end{itemize}

Hereafter, we call as ``CMB+BK15" the dataset using Planck 2018 likelihood combined with the Bicep/Keck experiment results. When BAO and SNe Ia data are considered, we indicate the dataset with ``CMB+BK15+Ext"
\begin{figure}[!t]
\centering
\includegraphics[scale=0.4]{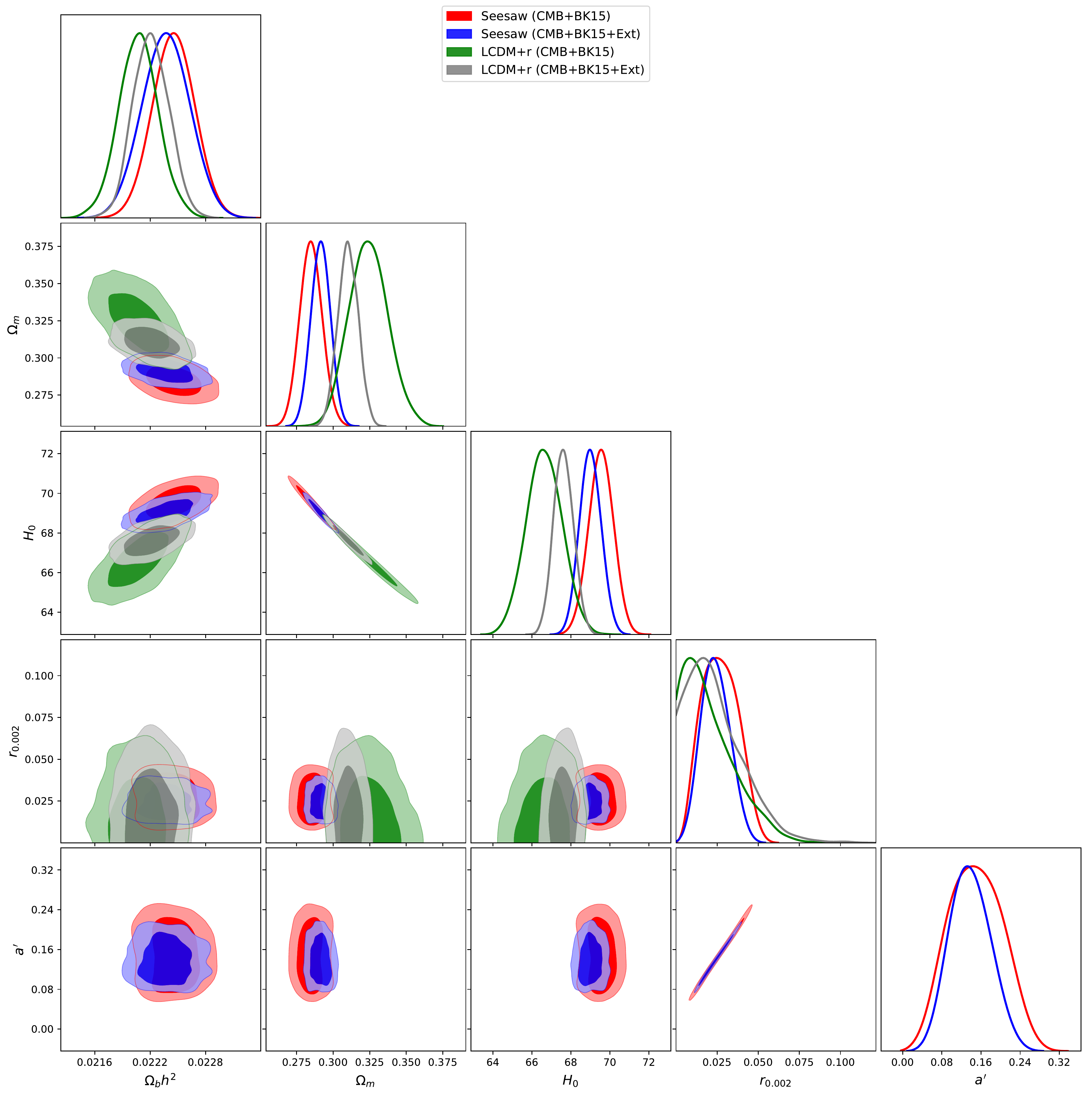}
\caption{2D C.L. and posterior distributions of SeeSaw and LCDM+r model, using the two dataset presented in the text.}
\label{Fig:Analysis}
\end{figure}
\section{Results}

The main results of our analysis are presented in figure \ref{Fig:Analysis}, where both $68\%$ and $95\%$ C.L. and posterior distributions are shown in comparison with those of the $\Lambda$CDM+$r$ model. We note that the SI model prefers higher values of the baryonic density, always compatible at $1\sigma$ with the standard cosmology. On the other hand, the constraints on the $\Omega_ch^2$ (and the total matter density, $\Omega_m$)  are significantly different for the two models. For example, for the SI model, we obtain $ \Omega_c h^2 = 0.1146 \pm  0.0012$ (CMB+BK15 data set) and $ \Omega_c h^2 = 0.1157 \pm 0.0009$ (CMB+BK15+Ext data set), while the $\Lambda$CDM+r analysis results furnish $ \Omega_c h^2 = 0.1212 \pm  0.0020$ (CMB+BK15 data set) and $ \Omega_c h^2 = 0.1189 \pm 0.0011$ (CMB+BK15+Ext data set). This in turn leads to a higher value of the  Hubble constant predicted by the SI model, i.e., $H_0= 69.54 \pm 0.53$ (CMB+BK15) and $H_0= 69.00 \pm 0.42$ (CMB+BK15+Ext), which alleviates the  well-known $H_0$-tension problem (see e.g. \cite{Verde:2019ivm,Alcaniz:2019kah} and references therein). For instance, considering the value obtained from CMB+BK15 data, the discrepancy between the SI model prediction and the latest measurement of $H_0$ using Cepheid  variables  and  type  Ia  Supernovae, $H_0 = 74.03 \pm 1.42$ km/s/Mpc~\cite{Riess:2019cxk}, is $\sim 2.97\sigma$ whereas if the value considered is the one obtained using geometric distance measurements to megamaser-hosting galaxies, $H_0 = 73.9 \pm 3.0$ km/s/Mpc~\cite{Pesce:2020xfe}, the discrepancy lowers to $\sim 1.43\sigma$.

The free parameter $a^\prime$ is constrained to be positive by more than $3\sigma$. We obtain $a^\prime = 0.15 \pm 0.05$ at $68\%$ C.L. for CMB+BK15 data set and $a^\prime = 0.14 \pm 0.03$ at $68\%$ C.L. for CMB+BK15+Ext. It is worth mentioning that the slow-roll analysis in section \ref{SR Analysis} indicates that null radiative corrections are compatible with Planck data  at $68\%$ CL (see figure \ref{Fig:ns_r}). A non-null value for $a^\prime$, therefore, could be a complementary probe for physics beyond the standard model. In particular, positive radiative corrections  favour type II seesaw mechanism, as discussed in section \ref{Seesaw Sec}. For $a^\prime = 0.15$, the amplitude of scalar perturbations in eq. (\ref{eq:PR}) impose the value $\lambda \simeq 3 \times 10^{-5}$ for the potential amplitude. 
We can use these estimates to constrain the couplings in radiative corrections $a$ using eq. (\ref{eq:alinha}) and, consequently, the Yukawa couplings for the neutrino sector, $h$ of eq.  (\ref{TypeIIradiative}), at inflationary energy scale. For CMB+BK15 data set we obtain $ 4.92 \times 10^{-7}\leq a \leq 1.90 \times 10^{-6}$ at $68\%$ (C.L.).

We emphasize that this is an important result since the Yukawa couplings are associated with the standard neutrinos masses through eq. (\ref{YukawaII}). In order to evaluate the impact of this bound at neutrino's masses we should use renormalization group equations to obtain $h^i$ at the electroweak scale and then consider the full flavor structure of the PMNS matrix \cite{Tanabashi:2018oca}. Such analysis is currently in progress and will be reported in a forthcoming communication.

\begin{figure}[!t]
\centering
\includegraphics[scale=1]{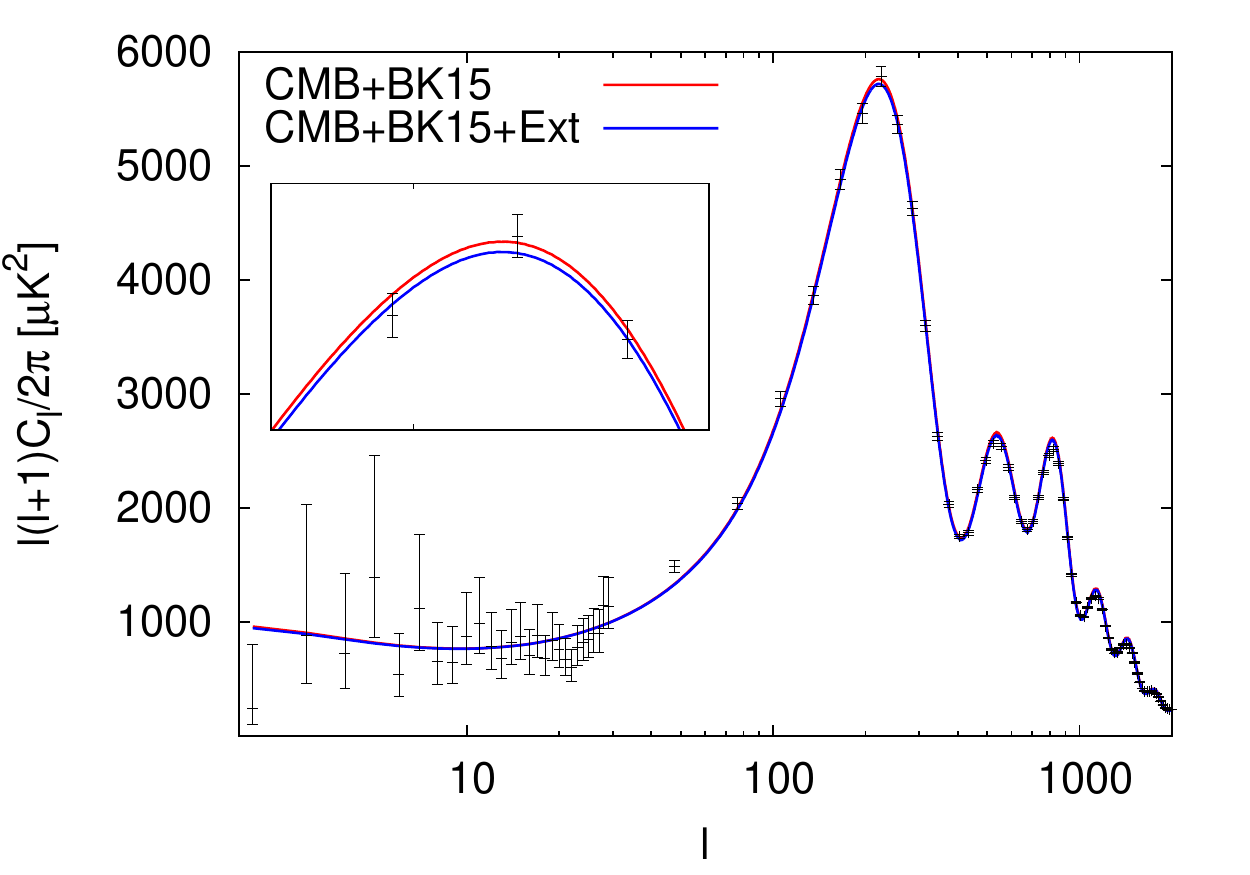}
\caption{Temperature power spectra obtained by using best fit values of our analysis on the SeeSaw inflation model in strong coupling regime, i.e. $\xi=100$. The two curves refer to analyzes with different data, namely Planck2018+BK15 (red line) and Planck2018+BK15+BAO+Pantheon (blue line). For comparison, the Planck data are also plotted. In the box, a zoom of the first anisotropy peak at $\ell \sim 200$.}
\label{Fig:TT_bestfit}
\end{figure}

Finally, we note a strong degeneracy between the $a^\prime$ and tensor-to-scalar ratio parameters, which leads to the narrow constraint $r_{0.002}=0.03 \pm 0.01$ (for the extended data set considered) excluding the absence of polarization modes in the CMB (i.e. primordial gravitational waves) at more than $3\sigma$. 
This is expected by the theory, since we foresee the probability distribution spread along the theoretical prediction of figure \ref{Fig:ns_r}. In particular, the slow-roll analysis of section \ref{SR Analysis} yields $r=0.026$ and $n_S=0.986$ for $a^\prime=0.15$. 
For completeness, we also show in figure \ref{Fig:TT_bestfit} the temperature anisotropy power spectrum obtained using the best fit values of our analysis. 
\section{Discussion and Conclusions}

Recently, growing interest has been given to inflationary scenarios with a non-minimal coupling between the inflaton field and the Ricci scalar, since this assumption may significantly modify the theoretical predictions of the simplest chaotic scenarios. Notwithstanding, this connection between the inflaton and gravity may give rise to a new scale of unitarity loss, affecting the predictive aspect of the model. Taking this into account, we connected the dynamics of a non-minimal inflationary scenario with neutrinos' physics, and obtained a unitarily safe and  predictive model. 

We investigated the observational viability of the model using a Boltzmann solver code in order to analyse the temperature anisotropy power spectrum in the light of the most recent CMB, BAO and SNe data. For that purpose, we fixed the non-minimal coupling $\xi$ to $100$ in order to obtain the best sensitivity for the observable $a^\prime$, as explained earlier. For the Hubble expansion, for instance, we obtained $H_0 = 69.54 \pm 0.53$ km/s/Mpc (with CMB+BK15 data) and $H_0 = 69.00 \pm 0.42$ km/s/Mpc (with CMB+BK15+Ext dataset), which alleviate the $H_0$-tension problem when compared with the $\Lambda$CDM+r predictions. 


The most interesting result obtained concerns the radiative sector ($a^\prime$). We found that the best-fit value for $a^\prime$ is positive constrained by more than $3\sigma$, favouring type II seesaw as the mechanism for generating neutrinos masses over type I. In particular, we obtain $a^\prime = 0.15 \pm 0.05$ at $68\%$ C.L. for CMB+BK15 data set. This is a remarkable result, since $a^\prime$ can be a complementary probe for physics beyond the Standard Model. We shall consider in a forthcoming communication the effects of this bound on a type II seesaw extension of the Standard Model.



%
%
\section*{Acknowledgments}
J.G.R. is supported by Conselho Nacional de Pesquisa e Desenvolvimento Cient\'ifico - CNPq. MB acknowledge Istituto Nazionale di Fisica Nucleare (INFN), sezione di Napoli, iniziativa specifica QGSKY. MC acknowledges the Programa de Capacita{\c c}{\~a}o Institucional (PCI-ON), supported from MCTIC/CNPq. JSA acknowledges support from CNPq (grant Nos. 310790/2014-0 and 400471/2014-0) and FAPERJ (grant No. E-26/203.024/2017). We also acknowledge the authors of the CosmoMC (A. Lewis) code and ModeCode (M. Mortonson, H. Peiris and R. Easther). This work was developed thanks to the High Performance Computing Center at the Universidade Federal do Rio Grande do Norte (NPAD/UFRN) and the National Observatory (ON) computational support. 

\bibliographystyle{JHEP}
\bibliography{bibliografia}

\end{document}